\documentclass[manuscript,screen]{acmart}

\AtBeginDocument{%
  }

\usepackage{graphicx}
\usepackage{algorithm}
\usepackage{algpseudocode}
\usepackage{subcaption}

\begin{document}

\title[Data-Driven Decision Making using a What-If Machine]{Should you make your decisions on a WhIM? Data-Driven Decision making using a What-If Machine for Evaluation of Hypothetical Scenarios}

\author{Jessica Maria Echterhoff}
\authornote{Work done at Microsoft.}
\email{jechterh@ucsd.edu}
\orcid{1234-5678-9012}
\affiliation{%
  \institution{University of California San Diego}
  \streetaddress{Gilman Dr}
  \city{San Diego}
  \state{California}
  \country{USA}
  \postcode{92037}
}
\author{Bhaskar Sen}
\author{Yifei Ren}
\author{Nikhil Gopal}
\email{\{bhaskar.sen,yifeiren,nikhil.gopal\}@microsoft.com}
\affiliation{%
  \institution{Microsoft}
  \streetaddress{P.O. Box 1212}
  \city{Dublin}
  \state{Ohio}
  \country{USA}
  \postcode{43017-6221}
}

\renewcommand{\shortauthors}{Echterhoff et al.}
\newcommand{\mycomment}[1]{}

\begin{abstract}
What-if analysis can be used as a process in data-driven decision making to inspect the behavior of a complex system under some given hypothesis. We propose a What-If Machine that creates hypothetical realities by resampling the data distribution and comparing it to the an alternate baseline to measure the impact on a target metric. Our What-If Machine enables both a method to confirm/reject manually developed intuitions of practitioners as well as give high-impact insights on a target metric automatically. This can support data-informed decision making by using historical data to infer future possibilities. Our method is not bound by a specific use-case and can be used on any tabular data. Compared to previous work, our work enables real-time analysis and gives insights into areas with high impact on the target metric automatically, moving beyond human intuitions to provide data-driven insights.
\end{abstract}

\maketitle

\section{Introduction}
In the realm of decision-making, uncertainty is an ever-present factor \cite{kale2019decision, boukhelifa2017data, greis2017designing}. Organizations and individuals alike are often faced with complex choices that can have far-reaching consequences. In this context, the technique of ``what-if'' hypothetical scenario analysis emerges as a valuable tool for navigating this uncertainty and aiding data-informed decisions. ``What-if'' analysis refers to the analysis of a possibility and measures its impact if it were to be implemented. Examples of what-if questions can be \textit{``What if we opened another branch at location Y, how would it impact our revenue?''\cite{bird2012assessing, golfarelli2006designing}; ``What if we invested in automatically detecting electricity outages using new sensors, would it reduce the number of customers we impact?}
By exploring various potential outcomes under different conditions in the data, this approach can provide a structured framework for evaluating options and mitigating risks \cite{hassani2016scenario}. What-if scenario analysis allows decision-makers to project their (potential) decisions into the future and assess how they might interact with other factors (like key performance indicators and target metrics). By considering a range of potential future states, organizations and individuals can refine their strategic plans, ensuring they are robust and adaptable in the face of changing circumstances. This leads to more agile decision-making and a higher degree of preparedness for various contingencies. Some of the reasons why data is recently treated as  ``a first-class citizen, on a par with code'' \cite{tae2019data}. Hypothetical scenario analysis presents different outcomes and their implications to stakeholders \cite{wollenberg2000using}. This process is often tightly connected with data science, described as a rational ``data-driven''  process of ``discovery'' that reveals the underlying nature of a domain \cite{kim2016emerging, rajan2013informatics, van2014data}. Data-driven hypothetical scenario analysis helps decision-makers identify the most efficient ways to allocate resources under different conditions and guide long-term strategic planning \cite{pettit2020data}. However, evaluating individual scenarios takes time and has its challenges, encouraging a data science team to focus on a single problem at-a-time \cite{muller2019data}. When one problem is tackled at a time, the process of evaluating hundreds of possible action becomes prohibitively time-consuming. 

Our work makes a \textbf{contribution} towards efficient data-driven decision making with our  \textbf{data-driven hypothetical scenario evaluation tool (``What-If Machine'')} to quickly validate practitioner's ideas and gain insights into potential action items. Beyond the improvement of time-consuming steps of data-driven decision-making, it provides a novel simulation-based algorithm for the evaluation of the entire data space of hypothetical scenarios automatically, to give insights into possibilities to drive change. Potential use-cases of our system are project management, investment decisions for different use cases such as application engineering, human resource planning, resource allocation strategies (for e.g. energy outage evaluation or climate change adaptation). 

\section{Background and Related Work}
\subsection{What-If Analysis in Data-Driven Decision-Making}
What-if analysis in the context of data-driven decision making measures what changes in a set of independent variables impact on a set of dependent variables with reference to a given simulation model \cite{philippakis1988structured}. This is tightly connected with human (data-driven) decision-making, which is associated with increased productivity for practitioners \cite{brynjolfsson2019data}. What-if analysis has shown to be helpful for decisions for which ``\textit{discoveries''} need to be made within data, and decisions that repeat, especially at scale, so decision making can benefit from even small increases in accuracy provided by data analysis \cite{dsrel}.  What-if analysis in data-driven decision making targets complex queries over data e.g. for forming a business strategy or looking for causal relationships in science \cite{deutch2013caravan}.  Alternative tools let practitioners test performance in hypothetical situations for very specific cases, e.g. for machine learning model analysis. For example, \citet{wexler2019if} analyze the importance of different data features, and visualize model behavior across multiple models and subsets of input data. Compared to \citet{wexler2019if}, our tool does not tackle machine learning model evaluation, but focuses on any use case that is based on tabular data where changes in the underlying data distribution (not the model) and its impact on a target metric are of interest to drive strategic decisions. Other previous work evaluates what-if questions on an individual basis \cite{gathani2022augmenting}, whereas we provide a tool that can answer what-if questions automated and at scale for data-rich industries. 

\subsection{Visualization and Simulation Tools}
For data driven decision-making, visualization plays an important role \cite{kelling2009data}. Some previous work let users share their hypotheses within the interface \cite{gotz2006interactive, shrinivasan2008supporting, stasko2007jigsaw} while exploring data, which can help people to recall their discoveries \cite{gotz2008characterizing, ragan2015characterizing}. However, some practitioners might have developed hypotheses based on their domain knowledge, and not within the exploratory data analysis. Manually evaluating a hypothesis can then require a full visual data exploration cycle \cite{yalccin2016cognitive}  or uncertainty-aware data analysis  with multiple steps such as acquiring data, manipulating data, reasoning, characterizing and presenting insights \cite{boukhelifa2017data} which may be prohibitively time-consuming to do for possibly hundreds of scenarios. This can lead to practitioners using heuristics for their decision-making instead \cite{METHLING2022100013}. Visualization tools facilitate exploratory data analysis for decision-making, but fall short at supporting hypothesis-based reasoning \cite{choi2019concept}. Our work provides a tool to quickly evaluate practitioners intuitions as well as provides insights into the most promising hypotheses with an automated analysis of a gamut of possible scenarios in the data.

Previous work tackles the design of simulation experiments and the validation of simulation models \cite{10, 12, 13}, but do not specifically use the simulation to support decision-making. Others limit themselves to particular use-cases (e.g. healthcare \cite{smith2020use,dunke2021simulation}). Alternative approaches try to derive predictive models from data automatically \cite{lingel2016poetics} to support decision-makers. We provide a generalizable tool that can work for any use case that is based on tabular data, and includes a target metric within (e.g. resource allocation, risk assessment, environmental impact analyses, infrastructure projects).



\begin{figure}[b]
    \centering    
    \begin{subfigure}{0.6\textwidth}
        \centering
        \includegraphics[width=\linewidth]{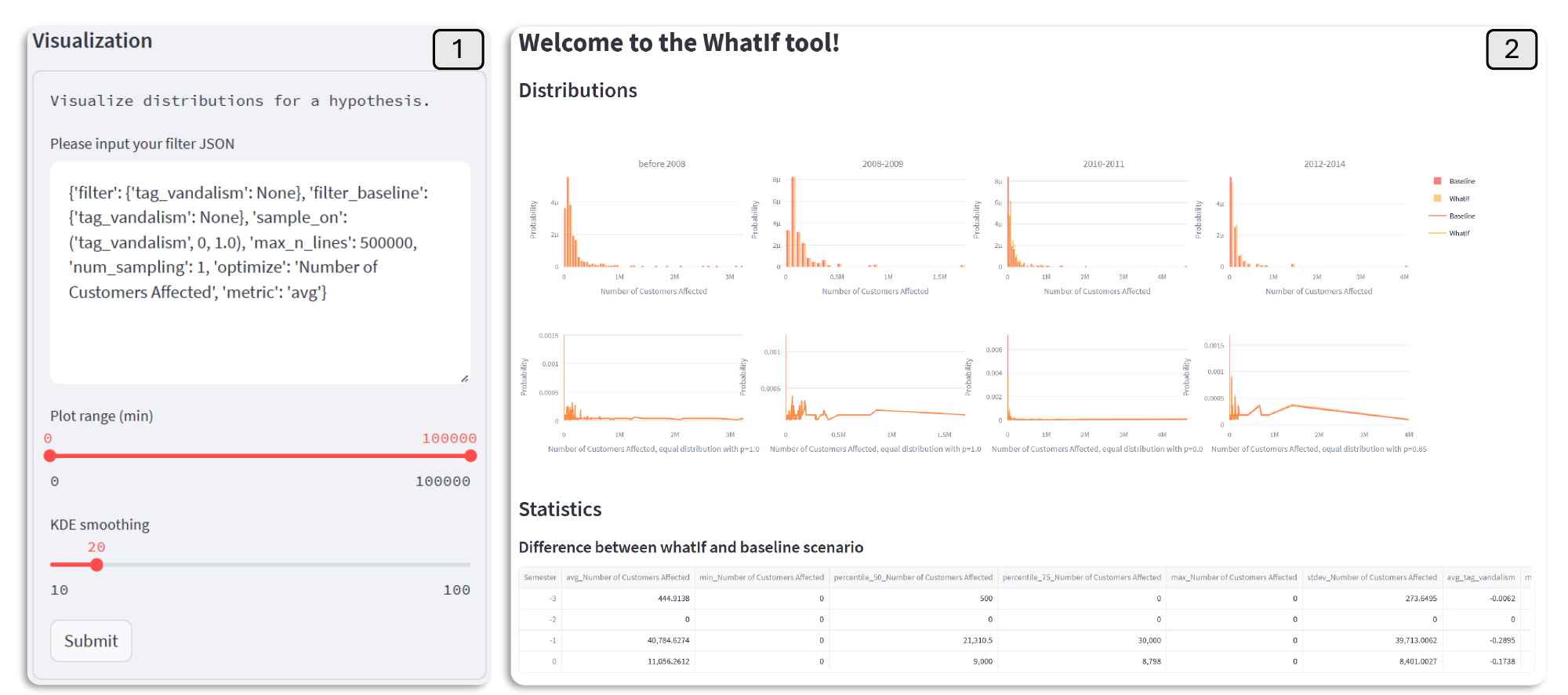}
        \label{subfig:a}
    \end{subfigure}
    \hfill
    \begin{subfigure}{0.39\textwidth}
        \centering
        \includegraphics[width=\linewidth]{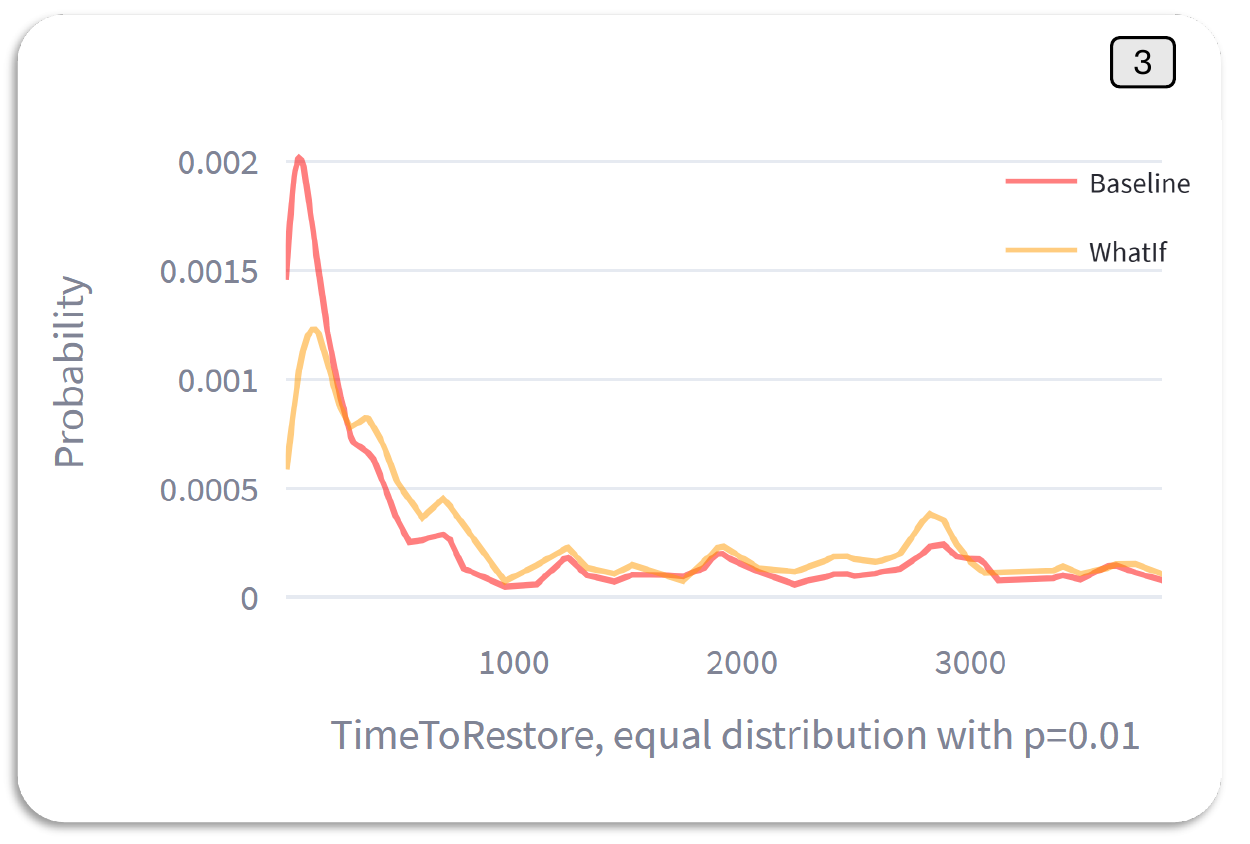}
        \label{subfig:ui2}
    \end{subfigure}
    \caption{User Interface of our What-If Machine for confirmative evaluation, (1) shows the control panel for setting a baseline and hypothetical scenario, as well as some hyperparameters such as graph smoothing and plot range. (2) shows the distribution visualization between the hypothetical and baseline scenario (e.g. if investing in avoiding electricity outages caused by vandalism can have a significant impact on the number of customers affected). It also shows common statistics and the potential gain of implementing the hypothetical scenario. (3) shows a close-up look of a what-if question -- namely if investing in weather resistant infrastructure can help to reduce the time to restore electricity after an outage. Reducing severe weather dependent outages seems to reduce the time to restore significantly, as indicated by the Komolgorov-Smirnoff Test p-value calculated for the two distributions.}
    \label{fig:ui}
\end{figure}

\section{Design of a What-If Machine}\label{sec:whatif}
\subsection{Design Implications}\label{sec:require}
On the one hand, there is a need for a tool for quickly confirming/disproving a hypothesis that was built with an experts domain knowledge with underlying data, as evidence-based decisions can improve organizational performance \cite{dremel2017audi}. On the other hand, developing an understanding of what problem should be solved in data science can be a complex and difficult process \cite{mao2019data}. Data scientists and decision makers such as program managers often make their decisions based on heuristics \cite{METHLING2022100013} or analysis of one scenario at a time \cite{muller2019data}. Practitioners might also get stuck in established thought patterns due to cognitive biases \cite{tversky1974judgment}. These insights sparked the initial idea of a tool to provide immediate feedback on impact of hypothetical scenarios, as well as providing promising the most possibilities itself. Based on the available literature, we develop the following design implications for a tool that should enable both quick evaluation of existing hypotheses (\emph{hypothesis confirmation/rejection}), as well as give a broad overview of impactful possibilities (\emph{hypothesis generation}).


\textbf{D1.} \textit{Hypothesis Confirmation/Rejection:} Previous work has shown that practitioners often seek to explicitly test their expectation against the data \cite{choi2019concept}. A What-If Machine should hence enable the user to quickly evaluate their own ideas and hypotheses and show the impacts and outcomes visually, to avoid going through the entire uncertainty-aware data analysis (acquire, manipulate, reason, characterize and present) \cite{boukhelifa2017data}. The tool should provide an analysis of impact (metrics), and present outcomes for scenarios. This supports the need for a quick way to analyze the potential effects of hypothetical changes. The tool should additionally allow for marginal evaluation to show the relationship of a variable to the target metric, and provide features that expedite the time-consuming aspects of evaluation. Overall, the tool should enable users to run simulations and projections to forecast potential outcomes based on historical data, saving time in manual analysis. 

\textbf{D2.} \textit{Hypothesis Generation:} The tool should provide the possibility to evaluate a multitude of hypothetical scenarios in a scalable manner. It should provide the possibility to browse scenarios for inspiration. It should give a quick overview on actions that can be taken as well as potential gains and stack-rank the outcomes based on impact, as previous work shows that weighing objectives and alternatives improves decisions \cite{METHLING2022100013}. The tool should provide visualizations that highlight a certain evidence magnitude, which can affect confidence in the decision-making \cite{peters2022confidence}. Prioritization and raking was found to be important for decision-confidence, as confidence reports are best explained by the difference between the posterior probabilities of the best and the next-best options, rather than by the posterior probability of the chosen (best) option alone, or by the overall uncertainty (entropy) of the posterior distribution \cite{li2020confidence}.


 \textbf{Both hypothesis confirmation/rejection and hypothesis generation rely on a similar underlying algorithmic idea: resample the historical data distribution using Monte-Carlo Sampling such that a hypothetical scenario is reflected, and report the impacts on a target metric.} We will describe the general idea of the procedure for an individual case of a manual hypothesis confirmation/rejection for \textbf{D1.} in Section \ref{sec:hc} and then scale it to evaluate the whole possible data space for \textbf{D2.} in Section \ref{sec:hg}.

\begin{figure}
    \centering    
    \begin{subfigure}{0.51\textwidth}
        \centering
        \includegraphics[width=\linewidth]{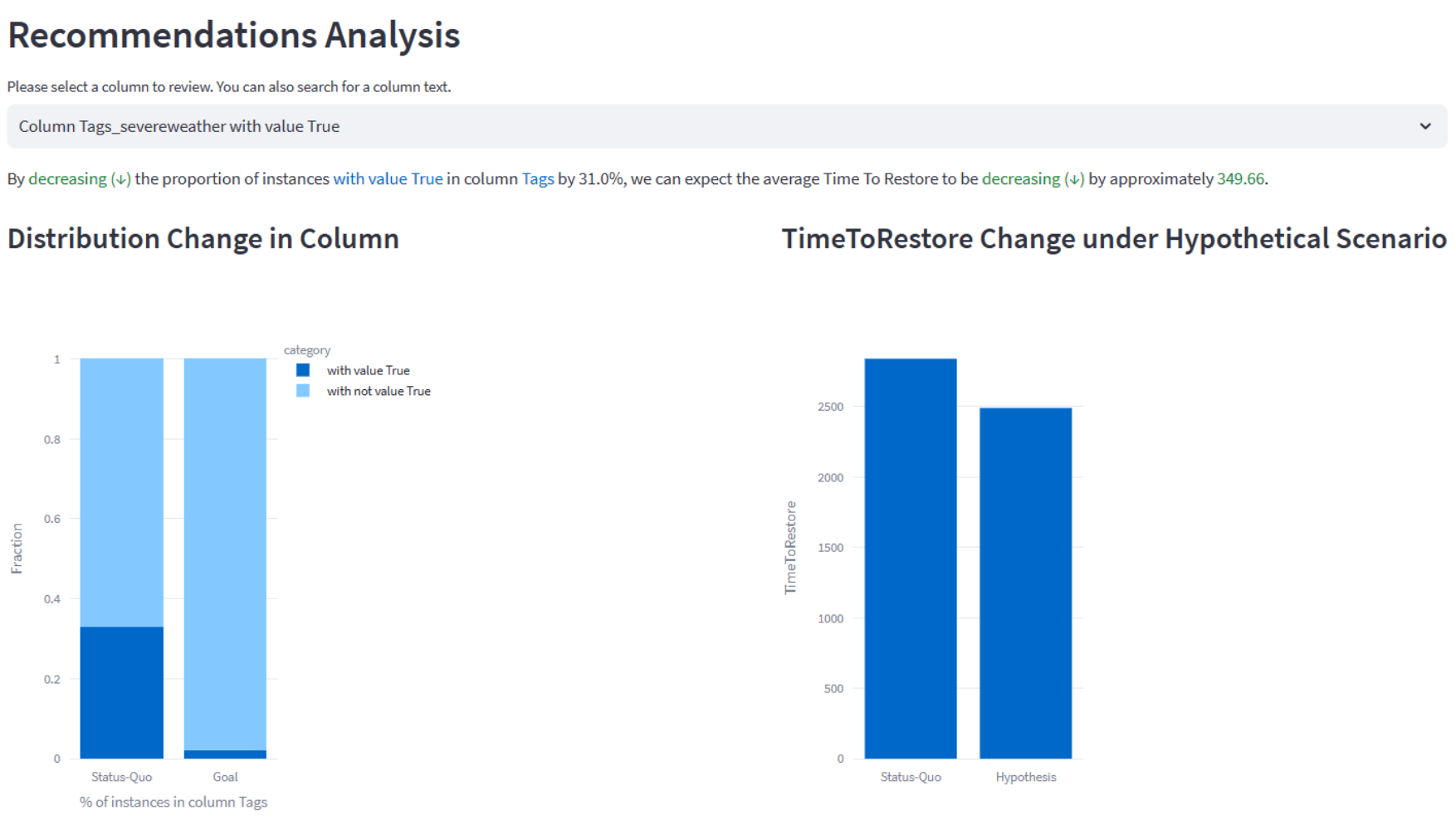}
        \label{subfig:a}
    \end{subfigure}
    \hfill
    \begin{subfigure}{0.48\textwidth}
        \centering
        \includegraphics[width=\linewidth]{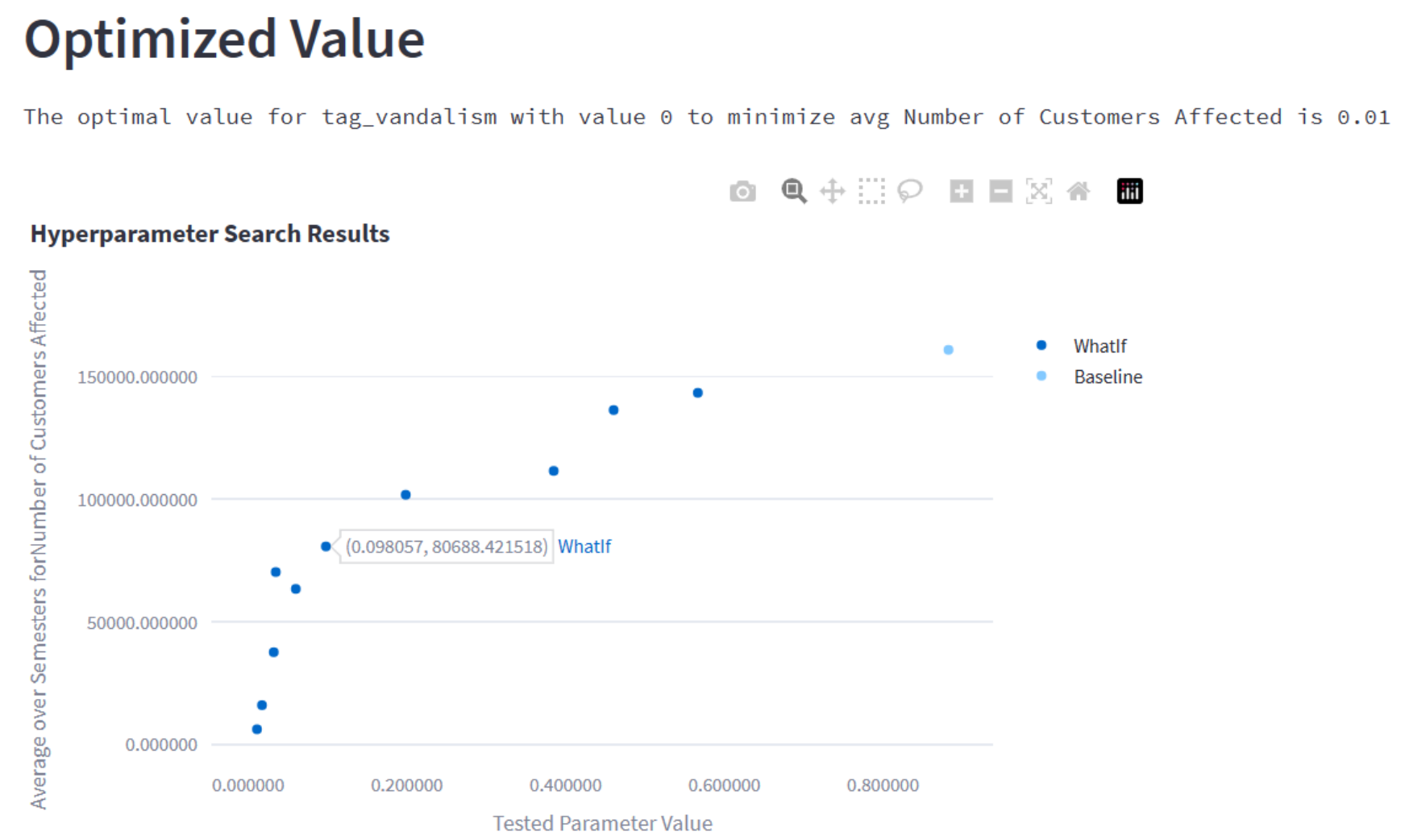}
        \label{subfig:ui2}
    \end{subfigure}
    \caption{Recommendation view, which shows a ranked list of potential action items that can be selected from a dropdown. When selected, it highlights the action to be done (e.g. reduce fraction of severe-weather outages from 32\% to 1\% , and potential benefit on the target metric (e.g. average time to restore will change from 2900 minutes to 2490 minutes(left). Right shows the marginal gain analysis, to determine the relationship between efforts and returns on the target metric. For example, here we show the relationship between electricity outages created by vandalism for different fractions of outages caused by vandalism in the data (e.g. 9.8\% of outages caused by vandalism), and average number of affected customers, to find out the benefit of investing in resiliency towards vandalism for electricity outages. }
    \label{fig:ui2}
\end{figure}

\subsection{Hypothesis Confirmation}\label{sec:hc}
To provide the ability to quickly evaluate the implications of a hypothetical scenario when a question arises, we implement a system that can sample a fraction of the data specified by the user and then compare it to a dedicated baseline specified by the user in real-time. To provide a visual indication on if the new (hypothetical) distribution is different, we show the two distributions (hypothetical scenario and baseline) as kernel density estimated distributions \cite{scott2015multivariate} (we are using the Silverman bandwidth \cite{silverman2018density}) as well as a binned histogram to show distributional changes in the two scenarios. In addition, the user can see standard statistical metrics such as average, percentiles, standard deviations etc., on their specified evaluation metrics. We show a visualization of the manual analysis component in Figure \ref{fig:ui}, for the what-if question ``What-if we invested in weather-resistant electricity infrastructure or reduce outages caused by vandalism, could we reduce the time to restore electricity after an outage and the number of customers affected?''. To get an overview of which data distribution might be optimal and look at the relationship between a target metric and a hypothetical change, the user can get a margin analysis of different potential distributions of a value in the data (Figure \ref{fig:ui2}). This marginal analysis enables an automatic evaluation of the optimal data distribution of a hypothetical scenario, to assist decision-makers in identifying the most efficient use of resources \cite{tsourapas2011evaluating}. The marginal analysis evaluates different hypothetical scenarios (e.g. fractions of severe-weather outages) and assess the target metric change to give an indication on if an investment may lead to linear, exponential or logarithmic gains. For this, we look for the optimal distribution to minimize our target metric $m$ (e.g. find the distribution of weather related outages for the target metric time to restore electricity) using Bayesian optimization. 
Formally, we are interested in 

\begin{equation}
x^* = \mathit{argmin}_x f(x)
\end{equation} 

Where for each evaluation metric $m$ and evaluation operator $P$
\begin{equation}
f (x) = P(m)
\end{equation}

For example, $P$ could be the mean or sum. Continuous values are mapped into bins to get an indication of which range of values might have an impact. Bayesian optimization \cite{frazier2018tutorial} uses a surrogate probabilistic model to approximate the objective function $f(x)$ and make predictions on unseen fractions. We use a Gaussian Process (GP) as the surrogate model with prior
\begin{equation}
 f(x) \sim GP(m(x), k(x, x'))
\end{equation}
where $m(x)$ is the mean function and $k(x, x')$ is the kernel (covariance) function that captures the correlation between points $x$ and $x'$.
Given observed data $D = {(x_1, y_1), (x_2, y_2), ..., (x_n, y_n)}$, where $y_i = f(x_i) + \epsilon$ and $\epsilon$ is the observation noise, the GP posterior is the updated GP model conditioned on the observed data. The acquisition function guides the selection of the next fraction. We use Expected Improvement (EI), which encourages exploration in regions where the model's performance is likely to improve significantly.

For optimization at each iteration $t$, given the current GP model and acquisition function, we select the next parameters to evaluate  the fraction of data value present in a given column,  denoted as $x_t$. We evaluate the objective function $f(x_t)$ and obtain the corresponding observed data $D = {(x_1, y_1), (x_2, y_2), ..., (x_t, y_t)}$, then update the GP model's posterior to incorporate the new data, obtaining the updated GP model. We repeat the process for a user predefined number of iterations (typically 10-20). 
 

\subsection{Hypothesis Generation}\label{sec:hg}
We use the aforementioned process to evaluate a single hypothesis evaluation at scale for hypothesis generation. Our idea is to use Monte-Carlo based sampling \cite{harrison2010introduction} when going over the entire potential search space. For example, if we have tabular data on different causes for electricity outages (e.g. severe weather, vandalism, infrastructure problems, technician error) and their impacts (e.g. number of customers affected or time to restore electricity), we can go through every data column and every unique data value, to systematically evaluate possible scenarios at scale. We apply the procedure from Section \ref{sec:hc} over all possible data columns and search for the optimal distribution of every unique data value (for categorical data) or data buckets of quantiles for numerical values. A detailed description of the process is shown as pseudocode in Algorithm \ref{alg:cap}. We ultimately display the most promising scenarios for display and review of the user, ranked by potential impact. We show a visualization in Figure \ref{fig:ui2}. 
\begin{algorithm}[t]
\caption{Pseudocode of the algorithm for automatic holistic what-if analysis over the data space $D$. Each (bucketed) unique value is resampled with an optimized fraction according to its effects on metric $m$, enabling us to automatically find the biggest effects caused by hypothetical change. whatif\_data returns the resampled data, that can then be compared e.g. to the full dataset.}\label{alg:cap}
\begin{algorithmic}
\Require $c \in D;n_{\mathit{sample}};n_{\mathit{unique}}; m$
\For{$c \in D$}
\If{$\mathit{len}(\mathit{unique}(c)) > n_{\mathit{unique}}$ and $\mathit{is\_num(c)}$}
\State $c = \mathit{bucket}(c)$ \Comment{Discretize numerical data} 
\EndIf
\For{$u_c \in  c$}
\State $x^{*} = \mathit{argmin_x} f(u_c, m, D)$ \Comment{Find optimal distribution of value $u_c$ in data $D$} 
\State $\mathit{whatif\_data} = \mathit{resample\_with\_fraction}(x^{*}, D, n_{\mathit{sampling}})$  \Comment{Resample data $n_{\mathit{sample}}$ times for robustness} 
\EndFor
\EndFor
\end{algorithmic}
\end{algorithm}

\subsection{Quantitative Evaluation}
\begin{table}[b]
  \caption{Quantitative evaluation of the What-If Machine. We are evaluating data from the past semester and their change of distribution over time to the following semester for each binary column in our proprietary dataset. We then use our sampling algorithm to recreate the distribution of the real change by sampling with the rate of change observed and compare the mean absolute error of all evaluation. Our results indicate that there is a 2-9\% error $(\pm 0.02)$. \label{1} }
  \label{tab:quant}
  \begin{tabular}{ll}
    \toprule
& MAE($\pm$ std)\\
\midrule
Semester –3 to -2  & $0.02 \pm (0.03)$\\
Semester -2 to -1 & $0.09 \pm (0.02)$\\
Semester -1 to 0 &	$0.06 \pm (0.02)$\\
  \bottomrule
\end{tabular}
\end{table}
We evaluate our tool based on historical data for real changes of variables observed in our (proprietary) databases over the timespan of two years. Quantitative evaluation of hypothetical scenarios based on real data is crucial to ensure accuracy of the decision-making base, and also help decision-makers to assess its results. By leveraging real-world data, we can construct real ``hypothetical'' scenarios, simulate the potential outcomes observed with our what-if tool and compare the simulated outcome to the real-world outcomes. By comparing the results of these evaluations with actual data from past events, decision-makers can validate the accuracy and reliability of our projections, enhancing the confidence in making informed choices for the future. We show the results of this evaluation in Table \ref{tab:quant}. 

\subsection{Usage Scenario}
We want to provide a usage scenario to show how our What-If Machine can help a program manager, Jamie, to develop and prioritize ideas and intuitions about their data. Let's consider Jamie's task to evaluate ways to reduce power outages in the USA and prioritize them for future planning. For this task, Jamie has access to power outage data between 2000 and 2014, which includes different causes for the outages as well as the impact on customers given by number of customers affected as well as time to restore electricity \cite{wirfs2014data}. Jamie is tasked to develop ideas to reduce the number of outages, time to restore electricity or customer impact. Jamie has an intuition that vandalism caused a decent number of outages and wants to evaluate to what extent investing into security systems will reduce the number of customers impacted. Jamie opens the What-If Machine and inputs his a hypothetical vandalism-caused outage rate of 0\% into the machine. Jamie finds that the impact on customers is not significant (Figure \ref{fig:ui}) in most timespans, indicated by the Komolgorov-Smirnoff test \cite{goodman1954kolmogorov} that compares the two distributions. This means that, assuming that security systems will decrease the probability of vandalism, the investment in security systems will not have much impact on the number of affected customers (Figure \ref{fig:ui}). Jamie checks what impact different fractions of vandalism have and if there might be diminishing returns when investing in security systems heavily. It seems like the number of customers affected is growing logarithmically with the percentage of outages caused by vandalism (Figure \ref{fig:ui2}). Now, Jamie has run out of ideas and would like to see if the tool can provide any ranked suggestions of investment areas, and finds that severe weather has an impact on the time to restore power, which could mean that if he increased the focus on making the infrastructure more resilient to weather conditions, customers could have their energy outages restored more quickly (Figure \ref{fig:ui2}). 

\section{Discussion}
\subsection{Design Walk-Through with Experts}
We showed practitioners and potential stakeholders a prototype of the What-If machine with the capabilities outlined in Section \ref{sec:whatif} within our organization, and got their informal feedback to prioritize further development efforts. We were interested in (1) if the tool would encourage to form what-if questions based on their own use cases and (2) which part of the what-if tool would be most impactful to practitioners. 9 out of 12 professionals started forming what-if questions within the feedback session, and everyone found applicable use-cases and data sources to be evaluated with our tool. Four people found the tool to be a great tool to move quickly and gage the potential of a scenario and all acknowledged utility of the tool in their decision-making process. One practitioner mentioned relationship between metrics and hypothetical evaluations, stressing that marginal gains are a super important question. Three practitioners especially liked the ranking capabilities of the automated analysis. Specifically the automated analysis provided by the tool was found to be interesting by two practitioners. It was acknowledged that the tool is giving insights into different angles and that what-if analysis is helpful to give recommendations to show the impact and drive the conversations for future strategies. 
Conversations with colleagues and practitioners also revealed need for more explanations on how to use the tool. This feedback lead us to implement a ``How to use this tool'' page. Colleagues also noticed that prioritizations could be more visible. Based on practitioners feedback we imagine implementing an ``auto highlighting'' procedure that automatically zooms into the largest deviations between the what-if and the baseline scenario in the future.

\subsection{Limitations and Future Work}
\subsubsection{Multi-Dimensional Analysis}
Our what-if tool can accurately resample changes in the data distribution to simulate hypothetical scenarios, as shown in our real-world evaluation in Table \ref{tab:quant}. We recognize that our algorithm primarily addresses the simulation of change of one variable within multiple possible hypothetical scenarios. While this is beneficial compared to previous work only evaluating one hypothetical scenario at a time \cite{gathani2022augmenting} as well as helping to provide a disentangled effect analysis of individual variable changes, future work could explore the ability to perform variations across multiple factors and evaluate cross-correlations to measure additional effects on the evaluation metrics.


\subsubsection{Finding the Abstraction Balance}
The evaluation of striking the right balance between offering comprehensive details and abstracting away complexity of the tool is a critical aspect that emerged from our design walk-through with experts especially when we talked to data scientists colleagues versus product managers. Different personas requested different amounts of detail that seemed to be contingent upon the extent of hands-on data involvement. Those who actively engage in extensive data manipulation and analysis regularly expressed a preference for in-depth elaboration that delved into the nuances of the tool's functionalities (e.g. statistical tests, correlations etc.). On the other hand, participants who approached the tool from a more streamlined utility perspective favored concise and quick evaluations that provided them with immediate insights. 

\subsubsection{Domain Knowledge Dependency}
Data science involves data, concepts, and methods, and thus requires expertise \cite{muller2019data} and ``good data won’t guarantee a good decision'' \cite{shah2012good}. A primary limitation of our study pertains to the persistence of domain knowledge requirement despite the automation achieved in our proposed framework. While we have successfully automated a crucial segment of the overall decision-making pipeline, it is important to acknowledge that the broader context of domain-specific understanding remains integral to the effective utilization of the tool. Our automated process significantly expedites certain aspects of the pipeline of high-level tasks for uncertainty-aware data analysis \cite{boukhelifa2017data} (e.g, acquire, manipulate and present), yet other stages necessitate human expertise for accurate decision-making and interpretation (e.g. reason why, characterize).  Our work predominantly addresses the "what" rather than the "why" of the given data. While our automation successfully provides answers to inquiries about factual information and patterns, e.g. to what extent severe weather incidents have an impact on customers affected, it does not provide causations or reasons behind the observed phenomena. Future endeavors can thus focus on incorporating mechanisms that further bridge the gap between automated procedures and human domain knowledge, ultimately striving for a comprehensive and integrated approach across the entirety of the pipeline. 


\section{Conclusion}

Our paper presents a versatile tool based on Bayesian Optimization and Monte-Carlo simulation that addresses the dynamic landscape of data-driven decision-making. Our ``What-If Machine'' enables quick data-driven hypothesis confirmation/rejection to speed up the data science pipeline as well as reveal potential high-impact areas automatically. By automating the process of generating ``what-if'' questions, our tool accelerates the exploration of various possibilities, providing real-time means of decision support. Simultaneously, the tool serves as an asset for practitioners seeking to evaluate their intuitions against data-driven insights, promoting a synergistic balance between human expertise and automated analytics. In sum, our tool introduces a novel approach to data-driven decision support, contributing to more informed and effective decision-making.


\bibliographystyle{ACM-Reference-Format}
\bibliography{sample-manuscript
}

\appendix

\end{document}